\documentclass[journal, 10pt]{IEEEtran}

\usepackage{amsmath,amsopn,amssymb,bm}
\usepackage{amsfonts}
\usepackage{amsthm} 
\usepackage{graphicx,xspace,color}
\usepackage{epsfig}
\usepackage{longtable,multirow}
\usepackage{array}
\usepackage{stfloats}
\usepackage{cuted}
\usepackage{cite}
\usepackage[nolist]{acronym}
\usepackage{soul}
\usepackage{algorithm}
\usepackage[noend]{algpseudocode}
\usepackage{algcompatible}
\usepackage{enumerate}
\usepackage{lettrine}
\usepackage{mathtools}
\usepackage{subcaption}

\newtheorem{lemma}{Lemma}
\newtheorem{definition}{Definition}
\newtheorem{theorem}{Theorem}

\usepackage[font=small,justification=centering]{caption}

\usepackage{booktabs}

\graphicspath{ {./figures/}}

\begin{document}

\begin{acronym}
	\acro{RU}{resource unit}
	\acro{RUs}{resource units}
	\acro{OFDMA}{orthogonal frequency division multiple access}
	\acro{OFDM}{Orthogonal frequency division multiplexing}
	\acro{SRA}{scheduling and resource allocation}
	\acro{WLANs}{wireless local area networks}
	\acro{AP}{access point}
	\acro{STA}{station}
	\acro{STAs}{stations}
	\acro{MCS}{modulation and coding scheme}
	\acro{UL}{uplink}
	\acro{TF}{trigger frame}
	\acro{TFs}{trigger frames}
	\acro{iid}{independent and indentically distributed} 
	\acro{SNR}{signal-to-noise ratio}
	\acro{PF}{proportional fairness}
	\acro{TWT}{target wake time}
	\acro{IoT}{Internet of Things}
\end{acronym}

\title{Optimal Resource Allocation in IEEE 802.11ax Uplink OFDMA with Scheduled Access}

\author{Konstantinos Dovelos and Boris Bellalta 

\thanks{\textcopyright~20XX IEEE. Personal use of this material is permitted. Permission from IEEE must be obtained for all other uses, in any current or future media, including reprinting/republishing this material for advertising or promotional purposes, creating new collective works, for resale or redistribution to servers or lists, or reuse of any copyrighted component of this work in other works.}
\thanks{K. Dovelos is with the Department
	of Information and Communication Technologies, Pompeu Fabra University, Barcelona (e-mail: konstantinos.dovelos@upf.edu).}
\thanks{B. Bellalta is with the Department
	of Engineering and Information and Communication Technologies, Pompeu Fabra University, Barcelona (e-mail: boris.bellalta@upf.edu).}
}

\maketitle

\begin{abstract}
We consider the scheduling and resource allocation problem in AP-initiated uplink OFDMA transmissions of IEEE 802.11ax networks. The uplink OFDMA resource allocation problem is known to be non-convex and difficult to solve in general. However, due to the special subcarrier allocation model of IEEE 802.11ax, the utility maximization problem involving the instantaneous rates of stations can be formulated as an assignment problem, and hence can be solved using the Hungarian method. In this paper, we address the more general problem of stochastic network utility maximization. Specifically, we maximize the utility of long-term average rates of stations subject to average rate and power constraints using Lyapunov optimization. The resulting resource allocation policies perform arbitrarily close to optimal and have polynomial time complexity. An important advantage of the proposed framework is that it can be used along with the target wake time mechanism of IEEE 802.11ax to provide guarantees on the average power consumption and/or achievable rates of stations whenever possible. Two key applications of such a design approach are power-constrained IoT networks and battery-powered sensor networks. We complement the theoretical study with computer simulations that evaluate our approach against other existing methods.
\end{abstract}

 \begin{IEEEkeywords}
 IEEE 802.11ax, OFDMA, Target Wake Time, Scheduling and Resource Allocation, Lyapunov Optimization
 \end{IEEEkeywords}

%%%%%%%%%%%%%%%%%%%%%%%%%%%%%%%%
\section{Introduction}
\lettrine{T}{HE} ever-growing demand for fast and ubiquitous wireless connectivity poses the challenge of delivering high data rates while efficiently managing the scarce radio resources. \ac{OFDM} has become a mainstream transmission method for broadband wireless systems, since it eliminates intersymbol interference by converting the frequency-selective channel into a set of flat-fading subchannels. Additionally, thanks to the independent fading of users' channels, efficient spectrum utilization can be attained by exploiting the so-called multiuser diversity. Specifically, \ac{OFDM} subchannels (i.e., subcarriers) can be dynamically allocated to multiple users according to their instantaneous channel conditions. The multiuser version of \ac{OFDM}, dubbed \ac{OFDMA}, has been therefore recognized as a key technology for next-generation wireless systems.

Towards this direction, the new IEEE 802.11ax amendment for high efficiency \ac{WLANs} combines multiuser multiple-input multiple-output (MU-MIMO) with \ac{OFDMA} in both downlink and \ac{UL} directions \cite{boris1}. Furthermore, in order to boost the \ac{UL} throughput, IEEE 802.11ax supports scheduled channel access along with the traditional contention-based access of WiFi via the use of \ac{TFs} and \ac{TWT} mechanism. More specifically, \ac{TFs} are used by the \ac{AP} to initiate multiuser \ac{UL} transmissions \cite{boris2}, whereas \ac{TWT} is used to schedule stations in time \cite{TWT}. These novel features of IEEE 802.11ax are expected to give WiFi a more reliable and cellular network-like performance. The efficiency of the scheduled \ac{OFDMA} transmissions, though, mainly depends on how the \ac{AP} selects the stations and allocates the available resources. Therefore, intelligent scheduling and resource allocation is crucial for attaining the best possible system performance.

\textbf{Scheduled Access in IEEE 802.11ax}. A IEEE 802.11ax \ac{AP} can solicit a group of stations to commence a multiuser \ac{UL} transmission by broadcasting a special control frame called \ac{TF}. The \ac{TF} contains the necessary information about the upcoming UL multiuser transmission, such as the list of selected stations and their transmission parameters, e.g., allocated RU, transmit power, and \ac{MCS} in the case of \ac{UL} \ac{OFDMA}. After receiving a \ac{TF}, the selected stations start to transmit synchronously for a given period of time. 

On the other hand, \ac{TWT} mechanism allows stations to agree with the \ac{AP} on a common \textit{wake time schedule}, namely, specific time instants to access the channel for either receiving or sending data; the rest of the time they remain in sleep mode. This enables stations to wake up only when required and minimize power consumption, a feature that is particularly important in \ac{IoT} and sensor networks. In addition to saving power on the station side, \ac{TWT}  reduces the contention level significantly, even supporting a collision-free and deterministic operation. For more details on \ac{TWT}, we refer the interested reader to \cite{TWT}. 

\section{Related Work and Contributions}
There is a plethora of literature studying the resource allocation problem in \ac{OFDMA}. However, most of those works assume that multiple  (possibly non-consecutive) subcarriers can be assigned to a single user. Under this assumption, it has been shown that greedy subcarrier allocation, i.e., each subcarrier is assigned to the user with the highest channel gain, along with waterfilling maximizes the sum-rate of downlink OFDMA systems \cite{SRM_OFDM}. Similar greedy allocation schemes have been derived for the weighted-sum rate maximization problem and the stochastic network utility maximization problem subject to minimum average rate constraints for both downlink \cite{Cioffi, Bill, Giannakis, Wang} and uplink \cite{ul_ofdma1},  \cite{ul_ofdma2} channels. None of these methods, though, can be readily applied to IEEE 802.11ax \ac{OFDMA} due to its peculiar subcarrier allocation model; in IEEE 802.11ax consecutive subcarriers are grouped into \ac{RUs}, and each user is assigned to one RU at most. 

There are few recent works on the scheduling and resource allocation problem for UL OFDMA in IEEE 802.11ax. In~\cite{Min_Padding}, the authors proposed a framework based on Lyapunov optimization to dynamically adjust the OFDMA transmission duration so that padding overhead is minimized. To do so, they assumed flat fading across the \ac{RUs} and considered fixed RU allocation in conjunction with round-robin user scheduling. The problem of joint user scheduling and RU allocation was firstly studied in \cite{UL_Schedulers1}, \cite{UL_Schedulers2}. Specifically, D. Bankov \textit{et al}. proposed a set of multiuser schedulers by formulating the unconstrained utility maximization problem as an assignment problem \cite{hungarian}. However, they focused on maximizing the utility of instantaneous user rates rather than their long-term average values. Therefore, their analysis cannot be readily used in scenarios where stations have requirements on the average power expenditure and/or average achievable rate.
\begin{figure*}[t]
	\centering
	\includegraphics[scale=0.62]{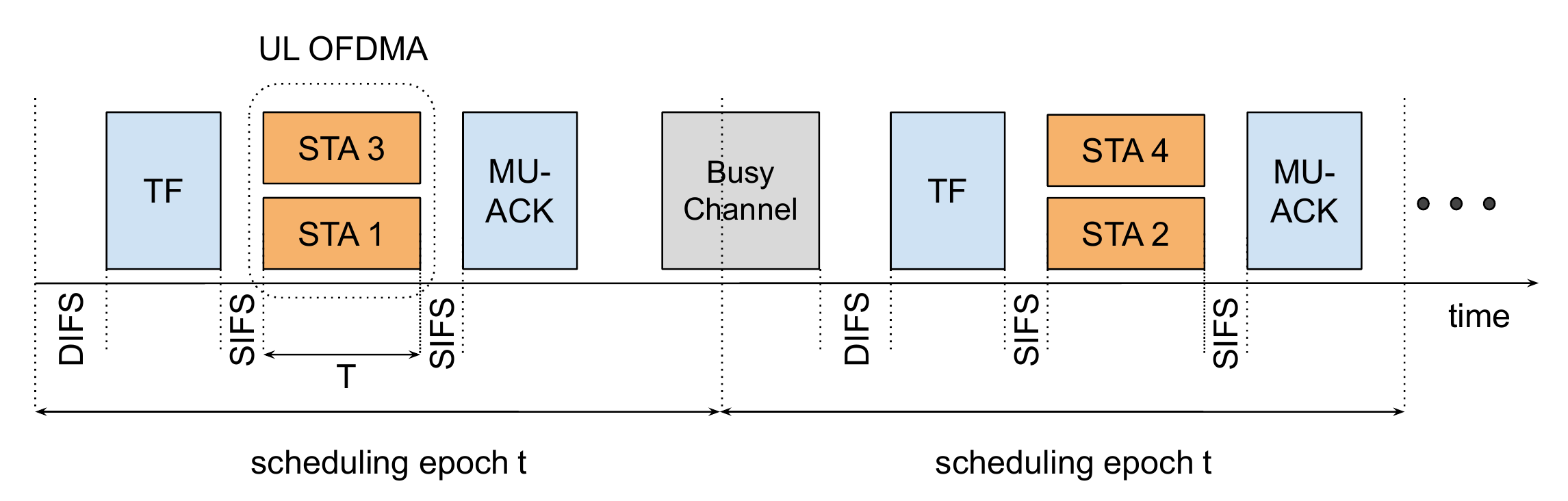}
	\caption{An example of UL OFDMA transmissions considered in the system model.}
	\label{fig1}
\end{figure*} 

In this paper, we fill this gap by addressing the stochastic utility maximization problem under average rate and power constraints. More particularly, we employ Lyapunov optimization techniques \cite{SNO1}, \cite{SNO2} and convert the stochastic optimization problem into a network stability problem of virtual queues. The latter problem is then solved by an online algorithm, known as drift-plus-penalty, whose performance is arbitrarily close to optimal. The main contributions of the paper are summarized as follows:
\begin{enumerate}[(i)]
	\item We address the problem of stochastic network utility maximization subject to average rate and power constraints for IEEE 802.11ax UL OFDMA. Specifically, we employ Lyapunov optimization to derive a near-optimal solution. Unlike previous works, our approach can be used along with \ac{TWT} to schedule stations in time and provide guarantees on their average power consumption and achievable rates whenever possible.
	\item We counter the case where the rate-constrained optimization problem is not feasible by means of weighted max-min (WMM) fair scheduling. Specifically, WMM method ensures minimum average rates for all stations whenever possible, and minimizes the relative constraint violation otherwise. 
	\item We show through simulations that the derived resource allocation policies outperform state-of-the-art methods such as proportional fairness in terms of average rates and average power consumption of stations.
\end{enumerate}

Throughout this paper, we use the following notation: $\mathcal{A}$ is a set; $\mathbf{a}$ is a vector; $\mathbb{E}[\cdot]$ denotes the expectation; $[\cdot]^+ = \max(\cdot,0)$; $\max(\mathcal{A})$ denotes the maximum element of set $\mathcal{A}$; $|\mathcal{A}|$ denotes the cardinality of set $\mathcal{A}$; and $[\mathbf{A}]_{ij}$ returns the $(i,j)$ entry of matrix $\mathbf{A}$.
 
\section{System Model}
Consider an uplink scenario where $K$ stations seek to communicate packets to the \ac{AP} of the network. We assume that the time axis is divided into scheduling epochs of equal duration, with epoch $t$ corresponding to the normalized time interval $[t, t+1)$. Henceforth, epoch and period will be used interchangeably. At the beginning of every epoch $t$, the AP initiates an UL OFDMA transmission of duration $T$, as shown in Fig. \ref{fig1}; if there is an on-going transmission, the  UL OFDMA transmission is deferred until the channel is sensed idle. The \ac{OFDMA} subcarriers are grouped into $N$ \ac{RUs}, and each RU consists of multiple consecutive subcarriers. Let $\mathbf{g}_{k,n}$ be the vector of channel gains of station $k$ on RU $n$. We collect all vectors $\mathbf{g}_{k,n}$ in the set $\mathcal{G}$. The \textit{channel state} in epoch $t$ is defined as $\mathcal{G}(t)$ and is assumed to evolve according to a block fading process. Hence, $\mathcal{G}(t)$ remains constant over epoch $t$ but is \ac{iid} over different scheduling epochs. We finally assume that the \ac{AP} has perfect knowledge of $\mathcal{G}(t)$ at the beginning of each epoch $t$.

\subsection{Resouce Allocation Actions}
Denote by $p_{k,n}(t)$ the transmit power of station $k$ on RU $n$ during the UL OFDMA transmission of epoch $t$. We assume that $p_{k,n}(t)$ takes values in a finite set $\mathcal{P}$ of available transmit powers. Moreover, let $s_{k,n}(t)$ be a binary variable that determines whether RU $n$ is assigned to station $k$ at period $t$ or not; $s_{k,n}(t) =1$ if RU $n$ is assigned to station $k$, and  $s_{k,n}(t) = 0$ otherwise. The \textit{resource allocation action} refers to the pair $(\mathbf{S}(t),\mathbf{P}(t))$, where matrices $\mathbf{S}(t)$ and $\mathbf{P}(t)$ are defined as $[\mathbf{P}(t))]_{kn} := p_{k,n}(t)$ and $[\mathbf{S}(t))]_{kn} := s_{k,n}(t)$ respectively. At the beginning of every epoch $t$, the \ac{AP} observes the random channel state $\mathcal{G}(t)$ and makes a resource allocation action $(\mathbf{S}(t),\mathbf{P}(t))$ within an option set $\mathcal{A}$. According to 802.11ax's RU model, RU assignment has to fulfil the following constraints:
\begin{align*}
\text{C1}: \ \ &\sum_{k=1}^K s_{k,n}(t)\leq 1, \ n=1,\dots,N,\\
\text{C2}: \ \ &\sum_{n=1}^N s_{k,n}(t) \leq 1, \ k=1,\dots, K.
\end{align*}
Constraint C1 ensures that users cannot share the same RU, while C2 ensures that every user is assigned to one RU at most. Let $\mathcal{F} =  \{0,1\}^{K\times N}\times \mathcal{P}^{K\times N}$. The option set $\mathcal{A}$ is then specified as
\[
\mathcal{A} = \left\{(\mathbf{S},\mathbf{P})\in \mathcal{F} \ | \ (\mathbf{S},\mathbf{P}) \ \text{fulfils} \ \text{C1 and C2} \right\}.
\]
 
 \subsection{Rate and Power Constraints}
 With channel realization $\mathbf{g}_{k,n}(t)$ and transmit power $p_{k,n}(t)$, user $k$ can transmit $r(p_{k,n}(t),\mathbf{g}_{k,n}(t))$ bits per OFDM symbol on RU $n$. The function $r(\cdot,\cdot)$ models the rate selection scheme, and has to conform with the 802.11ax restriction that a single MCS is employed over the subcarriers of a RU. Assume that there are $L$ MCSs, and let $\rho_l$ denote the bit rate of MCS $l$. If RU $n$ consists of $S_n$ data subcarriers and all subcarriers are used for transmission, then the set of achievable bit rates on RU $n$ is given by $\mathcal{R}_n = \{S_n\rho_1,\dots,S_n\rho_L\}$. The number of bits transmitted by user $k$ on RU $n$ during the scheduling period $t$ is denoted by $r_{k,n}(t)$ and is given by
\[
r_{k,n}(t) = r(p_{k,n}(t),\mathbf{g}_{k,n}(t))\frac{T}{T_{\textsc{ofdm}}},
\]
where $T_{\textsc{ofdm}}$ is the duration of an OFDM symbol, and $r(p_{k,n}(t),\mathbf{g}_{k,n}(t))\in\mathcal{R}_n$. Given the action $(\mathbf{S}(t),\mathbf{P}(t))$, the transmission rate and power consumption of user $k$ are calculated as
\begin{align*}
r_k(t) & := \sum_{n=1}^Ns_{k,n}(t)r_{k,n}(t),\\
p_k(t) & := \sum_{n=1}^N s_{k,n}(t)p_{k,n}(t).
\end{align*}
The time average expectations of $r_k(t)$ and $p_k(t)$ are defined, respectively, as
\begin{align*}
\bar{r}_k :&= \lim_{T\to\infty}\sup\frac{1}{T}\sum_{t=0}^{T-1}\mathbb{E}[r_k(t)],\\
\bar{p}_k :&= \lim_{T\to\infty}\sup\frac{1}{T}\sum_{t=0}^{T-1}\mathbb{E}[p_k(t)].
\end{align*}
We consider the case where each station $k$ has a minimum average rate requirement and a maximum average power expenditure limit specified by $r^{min}_{k}$ and $p^{max}_{k}$, respectively. Therefore, the resource allocation policy employed by the AP should ensure that $\bar{r}_k \geq r^{min}_{k}$ and $\bar{p}_k \leq r^{min}_{k}$, whenever doing so is possible. 
 
\section{Problem Statement}
Our objective is to design a resource allocation policy that selects $(\mathbf{S}(t),\mathbf{P}(t))$ at each scheduling epoch $t$ such that
\begin{equation}\label{eq:stochastic_problem}
\begin{matrix*}[l]
& \text{max}                & U(\bar{\mathbf{r}}) &\\\\
& \text{s.t.}       &  r^{min}_{k}\leq \bar{r}_k,  & k=1,\dots, K,\\[4pt]
& 					&  \bar{p}_k \leq p^{max}_{k}, & k = 1,\dots, K,\\[5pt]
&					& (\mathbf{S}(t),\mathbf{P}(t))\in\mathcal{A}, & t = 0,1,2\dots,
\end{matrix*}
\end{equation}
\\
where $\bar{\mathbf{r}} = (\bar{r}_1,\dots, \bar{r}_K)$, and $U(\cdot)$ is a concave, continuous and entrywise non-decreasing utility function. Due to the non-convexity of $\mathcal{A}$, the optimization problem under consideration is non-convex. We resort to Lyapunov optimization to derive a near-optimal solution. Next, we provide the definition of a so-called $O(\epsilon)$-optimal solution.

\begin{definition}
	Let $U^{opt}$ be the maximum utility of the problem defined in \eqref{eq:stochastic_problem}. A resource allocation policy is said to produce a $O(\epsilon)$-optimal solution if
	\begin{align*}
	U(\bar{\mathbf{r}}) &\geq U^{opt} - O(\epsilon)
	\end{align*}
	and all the constraints are satisfied.
\end{definition}

\section{General Solution via Lyapunov Optimization}
In this section, we use basic techniques of Lyapunov optimization, i.e., virtual queues and the drift-plus-penalty method, to solve the problem in \eqref{eq:stochastic_problem}.

\subsection{Virtual Queues}
In Lyapunov optimization, each time average constraint is associated with a virtual queue, and constraint satisfaction
is expressed as a queue stability problem. For each power constraint $\bar{p}_k\leq p^{max}_k$, consider a virtual queue that evolves over $t$ as
\[
Q_k(t+1) = \left[Q(t) - p_k^{max} + p_k(t)\right]^+, 
\]
where $Q_k$ denotes the queue backlog, the $p_{k}^{max}$ corresponds to the virtual constant service rate, and $p_k(t)$ is the virtual arrival process. Similarly for each rate constraint $r^{min}_{k}\leq \bar{r}_k$, consider a virtual queue with update equation 
\[
G_k(t+1) = \left[G_k(t) - r_k(t)+ r^{min}_{k}\right]^+.
\]
The following lemma establishes the connection between constraint satisfaction and queue stability.
\begin{lemma}
	If all queues $Q_k$ are mean rate stable, i.e., $\lim_{t\to\infty} \frac{\mathbb{E}[Q(t)]}{t} = 0$, then constraints $\bar{p}_k \leq p^{max}_{k}$ are satisfied~\cite{DPP_Convergence}.
	\begin{proof}
	See Appendix A.
	\end{proof}
\end{lemma}

\subsection{The Transformed Problem}
As in \cite{SNO1}, we transform the problem in \eqref{eq:stochastic_problem} into a form that involves only time averages rather than functions of time averages. Let $\bm{\gamma}(t) = (\gamma_1(t),\dots,\gamma_K(t))$ be a vector of \textit{auxiliary} variables chosen within a set $\Gamma$. The set $\Gamma$ must bound both the auxiliary and rate variables, and therefore is selected as
\[
\Gamma :=\{\bm{\gamma}\in\mathbb{R}^K \ | \  0 \leq \gamma_k \leq R_{max}, \ \forall k=1,\dots, K\},
\]
where $R_{max} = \max(\bigcup_{n=1}^N\mathcal{R}_n)$ is the maximum transmission rate over a RU. Now consider the following transformed problem:
\begin{equation}\label{eq:transformed_problem}
\begin{matrix*}[l]
& \text{max}                & \overline{U(\bm{\gamma})} &\\\\
& \text{s.t.}       &  \bar{\gamma}_k \leq \bar{r}_k,  & k=1,\dots, K,\\[4pt]
&         					    &  r^{min}_{k}\leq \bar{r}_k,           & k=1,\dots, K,\\[4pt]
& 					&  \bar{p}_k \leq p^{max}_{k}, & k = 1,\dots, K,\\[5pt]
& 					&  \bm{\gamma}(t) \in\Gamma, &  t = 0,1,2\dots,\\[5pt]
&					& (\mathbf{S}(t),\mathbf{P}(t))\in\mathcal{A}, & t = 0,1,2\dots,
\end{matrix*}
\end{equation}
\\
where $\overline{U(\bm{\gamma})} = \lim_{t\to\infty}\frac{1}{t}\sum_{\tau=0}^{t-1}\mathbb{E}[U(\bm{\gamma}(t))]$. The connection between \eqref{eq:stochastic_problem} and \eqref{eq:transformed_problem} is established as follows: consider a resource allocation policy $\pi$ which solves the problem defined in \eqref{eq:transformed_problem}. The maximum utility value $\overline{U(\bm{\gamma}_{\pi})}$ is then attained, and all constraints are satisfied, i.e., $\bar{\mathbf{r}}_{\pi} \geq \bar{\bm{\gamma}}_{\pi}$. Because $U(\cdot)$ is concave, it holds
\begin{equation}\label{eq:inequalities}
\bar{\mathbf{r}}_{\pi} \geq \bar{\bm{\gamma}}_{\pi} \Rightarrow U(\bar{\mathbf{r}}_{\pi}) \geq U(\bar{\bm{\gamma}}_{\pi}) \geq \overline{U(\bm{\gamma}_{\pi})}
\end{equation}
where the last inequality is Jensen's inequality for concave functions. What remains to show is that policy $\pi$ achieves a $O(\epsilon)$-optimal solution to the original problem in \eqref{eq:stochastic_problem}, namely $ \overline{U(\bm{\gamma}_{\pi})} \geq U^{opt} - O(\epsilon)$. If so, we have $U(\bar{\mathbf{r}}_{\pi}) \geq U^{opt} - O(\epsilon)$. 

\subsection{The Drift-Plus-Penalty Method}
Let $Q_k$, $Z_k$, and $G_k$ denote the backlogs of the virtual queues for constraints $\bar{p}_k \leq p^{max}_k$, $\bar{\gamma}_k \leq \bar{r}_k$,  and $r^{min}_{k}\leq \bar{r}_k$, respectively. The queue backlogs are updated according to the equations
\begin{align}\label{eq:virtual_queues}
Q_k(t+1) &= \left[Q_k(t) - p^{max}_k  + p_k(t)\right]^+\nonumber\\
Z_k(t+1) & = \left[Z_k(t) - r_k(t) + \gamma_k(t)\right]^+\\
G_k(t+1) &= \left[G_k(t) - r_k(t)+ r^{min}_{k}\right]^+\nonumber.
\end{align}
Let $\bm{\Theta}(t)$ denote the vector of all queue backlogs at epoch $t$, and consider the quadratic Lyapunov function
\[
L(\bm{\Theta}(t)) = \frac{1}{2}\sum_{k=1}^KQ^2_k(t) + \frac{1}{2}\sum_{k=1}^KZ^2_k(t) + \frac{1}{2}\sum_{k=1}^KG^2_k(t).
\]
The Lyapunov function is a scalar measure of network congestion. For example, if $L(\bm{\Theta}(t))$ is ``small", then all virtual queues are small, and if $L(\bm{\Theta}(t))$ is ``large", then at least one virtual queue is large. The (conditional) Lyapunov drift from period $t$ to the next one is defined as
\[
\Delta(\bm{\Theta}(t)) := \mathbb{E}[L(\bm{\Theta}(t+1)) - L(\bm{\Theta}(t)) | \bm{\Theta}(t)]. 
\]
\begin{algorithm}[H]
	\caption{Drift-plus-Penalty}
	
	\begin{algorithmic}[1]
		\Statex Set $\mathbf{Z}(0) = \mathbf{G}(0) = \mathbf{Q}(0) = 0$. For every scheduling period $t\in\{0,1,2,\dots\}$ do:
		\State Observe $\mathbf{\Theta}(t)$ and channel state $\mathcal{G}(t)$.
		\State Choose $\bm{\gamma}(t)\in\Gamma$ such that
		\[
		\max \ VU(\bm{\gamma}(t)) - \sum_{k=1}^K Z_k(t)\gamma_k(t)
		\]
		\State Choose $(\mathbf{S}(t),\mathbf{P}(t))\in\mathcal{A}$ such that
		\begin{align*}
		\max \ \sum_{k=1}^K (Z_k(t)r_k(t) &+ G_k(t)(r_k(t) - r_k^{min})\\
		&+ Q_k(t)(p^{max}_k-p_k(t))
		\end{align*}
		\State Update virtual queues according to \eqref{eq:virtual_queues}. 
	\end{algorithmic}\label{algo}
	\addtocounter{algorithm}{-1}
\end{algorithm}

If resource allocation actions are made at very period $t$ to greedily minimize $\Delta(\bm{\Theta}(t))$, then the queue backlogs are pushed towards a lower congestion state, which intuitively yields network stability that is equivalent to satisfying the desired time average constraints. To take into account the objective function as well, we consider the \textit{drift-plus-penalty} expression  $\Delta(\bm{\Theta}(t)) - V\mathbb{E}[U(\bm{\gamma}(t)) | \bm{\Theta}(t)]$, where $V>0$ is a control parameter whose role will be revealed later. One can show that the drift-plus-penalty expression is bounded as 
\begin{align}\label{eq:drift}
\Delta(\bm{\Theta}(t)) - V\mathbb{E}[U(\bm{\gamma}(t)) &| \bm{\Theta}(t)] \leq B - V\mathbb{E}[U(\bm{\gamma}(t)) | \bm{\Theta}(t)]\nonumber\\
&+ \sum_{k=1}^K Q_k(t)\mathbb{E}[p_k(t) - p^{max}_k | \bm{\Theta}(t)]  \nonumber\\
&+ \sum_{k=1}^K Z_k(t)\mathbb{E}[\gamma_k(t) - r_k(t) | \bm{\Theta}(t)] \nonumber \\
&+ \sum_{k=1}^K G_k(t)\mathbb{E}[r_k^{min} - r_k(t) | \bm{\Theta}(t)],
\end{align}
where $B = \frac{1}{2}\sum_{k=1}^K\left((p^{max}_k)^2 + (r^{min}_k)^2 + (P_{max})^2 + 3R^2_{\max}\right)$, $P_{max} = \max(\mathcal{P})$ denotes the maximum transmit power level, and $R_{max}$ is the maximum transmission rate over a RU. Please refer to Appendix B for the proof of the bound. The drift-plus-penalty (DPP) algorithm \textit{opportunistically} minimizes the right-hand side of \eqref{eq:drift}. The detailed steps of DPP are given in Algorithm \ref{algo}. The following theorem establishes the near-optimal performance of DPP.

\begin{theorem}
	Suppose the problem in \eqref{eq:stochastic_problem} is feasible and $\{\mathcal{G}(t)\}_{t=0}^{\infty}$ are \ac{iid} over the scheduling periods. Then for a given constant $V>0$, the drift-plus-penalty algorithm achieves an $O(1/V)$-optimal solution to \eqref{eq:stochastic_problem}.
	
	\begin{proof}
		See Appendix C. 
	\end{proof}
\end{theorem}

\section{Instantaneous Maximization Subproblems}
According to Theorem 1, a solution to the stochastic problem in \eqref{eq:stochastic_problem} can be obtained by solving a set of deterministic subproblems at every scheduling epoch $t$. The first one regards the auxiliary viarables used in transformed problem. Let $U(\bm{\gamma}(t)) = \sum_{k=1}^K U_k(\gamma_k(t))$. We then have
\begin{equation*}
\begin{aligned}
&\max
&&  \sum_{k=1}^K\left(V U_k(\gamma_k(t)) - Z_k(t)\gamma_k(t)\right)\\
&\ \  \text{s.t.}
&& \bm{\gamma}(t)\in \Gamma,
\end{aligned}
\end{equation*}
\\
The optimal auxiliary variables are given by
\[
\gamma_k^*(t) = \arg\max_{0\leq\gamma_k(t)\leq R_{max}} \left(V U_k(\gamma_k(t)) - Z_k(t)\gamma_k(t)\right).
\]
Next, the maximization problem for the resource allocation action $(\mathbf{S}(t),\mathbf{P}(t))$ is cast as 
\begin{equation}\label{eq:insta_max}
\begin{aligned}
&\max
&& \sum_{k=1}^K\sum_{n=1}^Ns_{k,n}(t)\phi_{k,n}(p_{k,n}(t))\\
& \ \ \text{s.t.}
&& s_{k,n}(t) \in\{0,1\}, \ \forall k, \forall n \\
&&& \sum_{k=1}^K s_{k,n}(t) \leq 1,\ \forall n\\
&&& \sum_{n=1}^N s_{k,n}(t) \leq 1, \ \forall k\\
&&& p_{k,n}(t) \in\mathcal{P},  \ \forall k, \forall n,
\end{aligned}
\end{equation}
where
\begin{align*}
\phi_{k,n}(p_{k,n}(t)) = Z_k(t) r_{k,n}(t) &+ G_k(t)(r_{k,n}(t) - r_k^{min}) \\
&+ Q_k(t)(p^{max}_k- p_{k,n}(t)).
\end{align*}
Since powers $p_{k,n}$ are independent of each other, we can first determine the optimal power allocation over each RU $n$ by solving
\[
\phi_{k,n}^*(t)  = \max_{p_{k,n}(t)\in\mathcal{P}}	\phi_{k,n}(p_{k,n}(t)).
\]
Once $\{\phi_{k,n}^*(t)\}_{k,n}$ are obtained, problem in \eqref{eq:insta_max} becomes
\begin{equation}\label{assign_problem}
	\begin{aligned}
	&\max
	&& \sum_{n=1}^N\sum_{k=1}^K  s_{k,n}(t)\phi^*_{k,n}(t)\\
	& \ \ \text{s.t.}
	&& s_{k,n}(t) \in\{0,1\}, \ \ \forall k,\forall n\\
	&&& \sum_{k=1}^K s_{k,n}(t) \leq 1, \ \ \forall n\\
	&&& \sum_{n=1}^N s_{k,n}(t) \leq 1, \ \ \forall k.
	\end{aligned}
\end{equation}
The above problem determines the optimal user/RU assignment for the scheduling period $t$, and is a classic assignment problem. Therefore, it can be optimally solved in $O(\max(K,N)^3)$ via the Hungarian method \cite{hungarian}. The overall computational complexity of the DPP algorithm is given by $O(K + KNML + \max(K,N)^3)$, where $M = |\mathcal{P}|$ is the number of available transmit powers and $L$ is the number of MCSs.

\section{Weighted Max-Min Fairness}\label{sec:wmm}
In this section, we present a practical algorithm to deal with the case where the problem in  \eqref{eq:stochastic_problem} is not feasible. To this end, we incorporate the average rate constraints into the utility function and formulate the weighted max-min problem
\begin{equation}\label{eq:max_min}
\begin{aligned}
&\max
&&  \min_{k}\left\{\frac{\bar{r}_k}{r_k^{min}}\right\}\\ \\
& \ \  \text{s.t.} 
&& \bar{p}_k \leq p^{max}_{k}, \ \ \forall k\\
&&&(\mathbf{S}(t),\mathbf{P}(t))\in\mathcal{A}, \ \ \forall t.
\end{aligned}
\end{equation}
Define the maximum relative constraint violation as

\[
\Delta : = \max_k\left\{\frac{r_k^{min} - \bar{r}_k}{r_k^{min}}\right\} = \min_k\left\{\frac{ \bar{r}_k}{r_k^{min}}\right\}. 
\] 
\\
The policy that solves \eqref{eq:max_min} ensures the desired minimum average rates of stations whenever possible, and minimizes the maximum relative constraint violation $\Delta$ otherwise. To solve the weighted max-min problem, we set $\gamma_k = r_k/r_k^{min}$ and consider a virtual queue with update equation
\[
 Z_k(t+1) = \left[Z_k(t) - \frac{r_k(t)}{r_k^{min}} + \gamma_k(t)\right]^+.
\]
Next, the optimal auxiliary variables are obtained by observing that
\begin{align}\label{eq:aux_ineq}
\max_{\bm{\gamma}\in\Gamma} \left(V\min_k\{\gamma_k\} - \sum_{k=1}^KZ_k\gamma_k \right) \nonumber\\
\leq \max_{\bm{\gamma}\in\Gamma} \left(\gamma_{min} \left(V-\sum_{k=1}^KZ_k\right) \right),
\end{align}
\\
where the time index is neglected for ease of notation, and $\gamma_{min} = \min_k\{\gamma_k\}$. According to \eqref{eq:aux_ineq}, the right-hand side expression is maximized by setting $\gamma_{min} = R_{max}$, i.e., choosing the maximum value for all auxiliary variables when $V > \sum_{k} Z_k$, and $\bm{\gamma}=\mathbf{0}$ otherwise. Therefore, we have
\begin{equation*}
\gamma^*_{k} = \left\{
\begin{array}{ll}
R_{max}, & \text{if} \ V > \sum_{k=1}^KZ_k \\ \\
0, &\text{otherwise}\\ 
\end{array}
\right. .
\end{equation*}
Finally, the optimal resource allocation action $(\mathbf{S}(t),\mathbf{P}(t))$ is determined as previously.

\section{Performance Evaluation}
In this section, we assess the performance of the proposed framework over different network conditions and average rate/power requirements. We simulate a IEEE 802.11ax WLAN with users following a full buffer traffic model, i.e., they always have data for transmission. The overheads associated with \ac{UL} \ac{OFDMA} transmissions are neglected, since they only scale the achievable rates of each resource allocation algorithm, and hence do not affect the comparison of the different algorithms. As in \cite{UL_Schedulers2}, we consider the following baseline schemes:
\begin{itemize}
	\item[-] \textbf{Sum-Rate Maximization (SRM):} in every epoch $t$, the scheduler chooses the stations who maximize the instantanous sum-rate. 
	\item[-] \textbf{Proportional Fairness (PF):} in every epoch $t$, the scheduler selects the stations who maximize the instantaneous weighted sum-rate, where the weight associated with station $k$ equals the inverse of the exponential moving average of its rate \cite{Fairness}. 
	\item[-] \textbf{Random Selection (RND):} in every epoch $t$, the scheduler uniformly and at random selects a set of stations. 
\end{itemize} 

\begin{figure*}
	\centering
	\begin{subfigure}{1.\columnwidth}
	\includegraphics[width=\columnwidth]{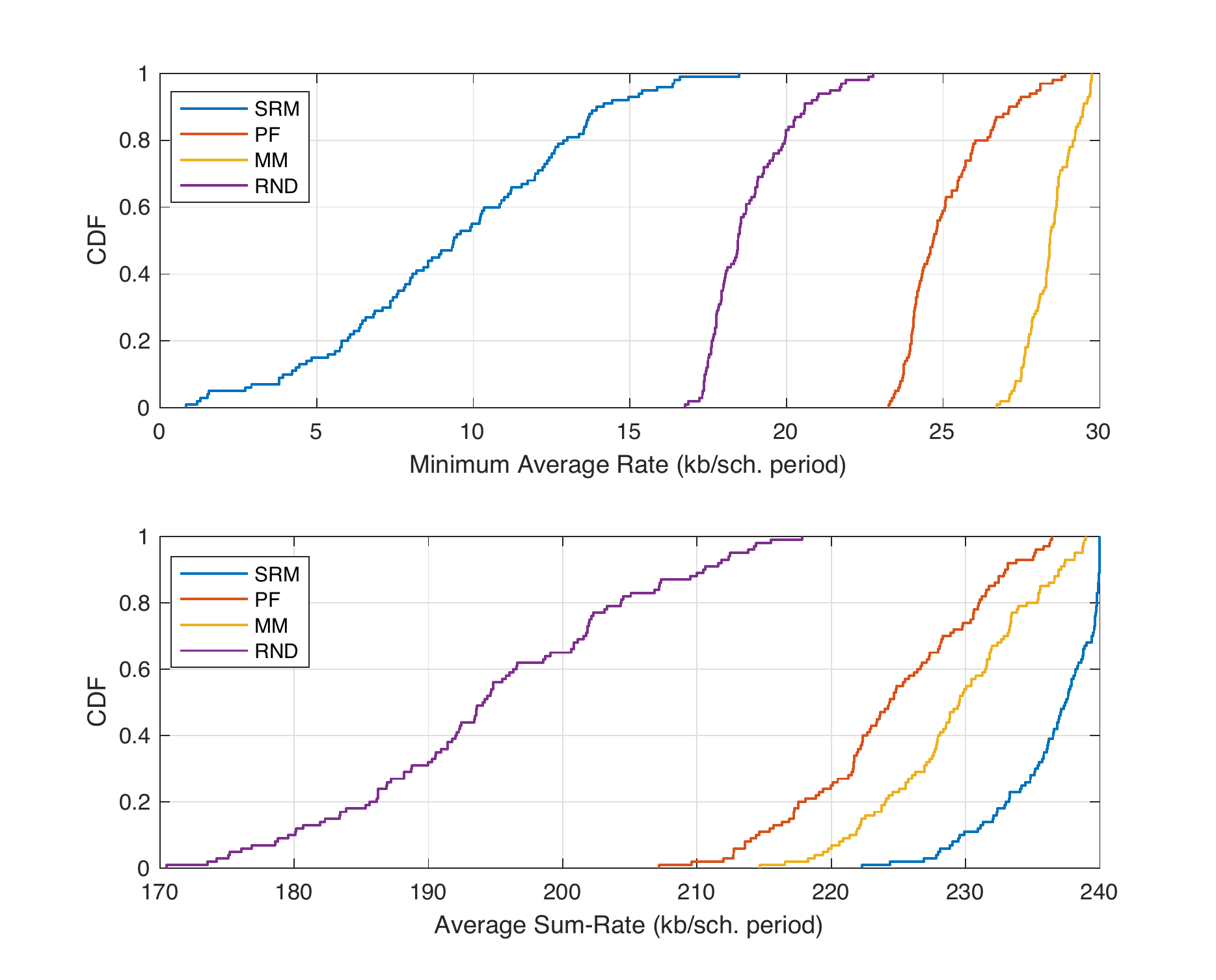}
	\caption{}
	\label{Fig:Result_1}
 	\end{subfigure}\hfil
	\begin{subfigure}{1.\columnwidth}
			\includegraphics[width=\columnwidth]{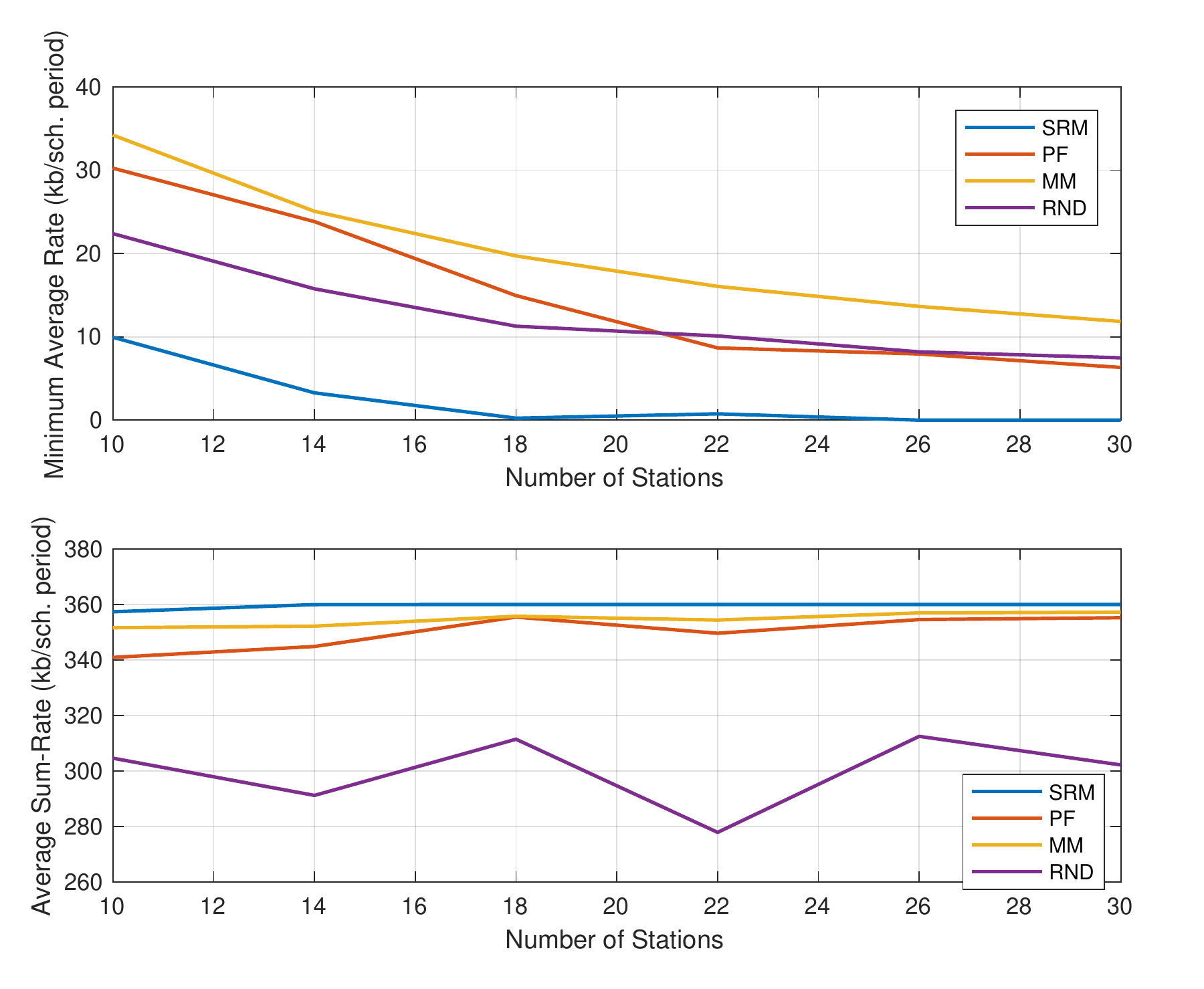}
			\caption{}
		\label{Fig:Result_2}
	\end{subfigure}
\caption{Results for unconstrained utility maximization: (a) empirical CDF of the minimum average rate of stations and the average sum-rate \\for a WLAN with $K = 8$ stations; (b) minimum average rate and average sum-rate vs. number of stations.}
\end{figure*}

\begin{table}[H]
	\centering
	\caption{Main simulation parameters.}\label{Table:sim_params}
	\begin{tabular}{ c | l | c} 
		\hline
		\textbf{Notation} & \textbf{Description}  & \textbf{Value}\\ 
		\hline
		$N$ & Number of RUs & $9$  \\
		$N_{sc}$ & Number of data subcarriers per RU & $24$ \\
		$\text{PL}_0$ & Pathloss at $1$ m & $20$ dB \\
		$a$ & Pathloss exponent & $4.4$  \\
		$d_{max}$ & Radius of the WLAN area  & $15$ m  \\
		$P_{max}$ & Maximum transmit power & $20$ dBm \\
		$T_{\textsc{ofdm}}$ & Duration of OFDM symbol & $16$ $\mu\text{s}$ \\
		$T$ & Duration of UL OFDMA transmission & $3.2$ $\text{ms}$ \\
		$V$ & Control parameter of DPP algorithm & $100$\\
		\hline
	\end{tabular}
\end{table}

\subsection{Simulation Model and Parameters}
The area of the WLAN is modeled by a circle of radius $d_{max}$ meters. The \ac{AP} is located at the center of the circle, and stations are uniformly distributed inside the circle with minimum distance from the AP of $1$ meter. Path attenuation is given by the log-distance model 
\[
\text{PL}_k = \text{PL}_0 + 10a\log_{10}(d_k) \quad (\text{dB})
\]
where $\text{PL}_0$ is the loss at the reference distance of 1m and $a$ is the pathloss exponent. The channel bandwidth is $20$ MHz and is divided into $N = 9$ \ac{RUs}. Each RU $n$ consists of $S = 24$ data subcarriers. We assume Rayleigh fading across the \ac{RUs}. Let $g_{k,n}$ denote the channel gain of station $k$ over RU $n$. Power $p_{k,n}$ is uniformly distributed among the subcarriers of RU $n$, and therefore the received \ac{SNR} at each subcarrier is given by
\[
\text{SNR}_{k,n}(p_{k,n})=  10\log_{10}\left(\frac{p_{k,n}}{S} \right) - \text{PL}_k + 10\log_{10}(g_{k,n})  \ (\text{dBm}).
\]
Based on the received \ac{SNR}, the maximal \ac{MCS} is selected, which is denoted by $l^*$. The bit rate and \ac{SNR} threshold of each \ac{MCS} are given in Table \ref{table: MCS} of Appendix D. The rate of station $k$ over RU $n$ is calculated as
\[
r_{k,n} = S \rho_{l^*} \frac{T}{T_{\textsc{ofdm}}}   \quad (\text{bits}/\text{sch. period}).
\]
The values of the main simulation parameters are given in Table \ref{Table:sim_params}. For the auxiliary variables, we consider the option set $\Gamma = \{S \rho_1,\dots,S\rho_L\}$; $\rho_1$ is the bit rate of BPSK with code rate $1/2$, and $\rho_L$ is the bit rate of $256$-QAM with code rate $5/6$.

\subsection{Results for Unconstrained Utility Maximization}
Consider the unconstrainted maximization problem $\max U(\bar{\mathbf{r}})$, where all stations transmit at the maximum power level $P_{max} = 20$ dBm. We implement max-min (MM) fairness, which is given by the utility function $U(\bar{\mathbf{r}}) = \min\{\bar{r}_k\}$. In the first experiment, we generate $1000$ network topologies, and for each network topology we run the scheduling algorithms for $10.000$ periods. Next, we average the rates of stations over all network topologies and calculate the empirical cumulative distribution function (CDF) of the minimum average rate and the average sum-rate. We are particularly interested in the minimum rate of stations since all stations will have to be served with high rates in next-generation WLANs.

Fig. \ref{Fig:Result_1} shows that MM policy provides a more fair rate allocation (in terms of minimum average rate) among stations compared to other policies. Furthermore, we see that maximizing the minimum rate of stations does not lead to significant loss in the sum-rate, as MM policy can achieve higher sum-rate than the RND and PF methods. On the other hand, SRM policy attains by definition the highest sum-rate but presents poor performance in terms of fairness. This is because SRM schedules the stations with favourable channel conditions, i.e., the ones close to the \ac{AP}, to achieve high throughput. However, selecting the stations with the best channel quality leads to starvation of other stations. Next, we assess the performance of each scheduling policy as the number of stations in the WLAN increases. Specifically, for a given number of stations, we generate a random network topology and run every scheduling policies for $10.000$ periods. The results in Fig. \ref{Fig:Result_2} show that SRM policy excludes the stations who are located far from the AP as the number of stations increases. On the other hand, MM policy serves all stations and delivers the highest average minimum rate. 
\begin{figure*}
	\centering
	\begin{subfigure}{0.97\columnwidth}
		\includegraphics[width=\columnwidth]{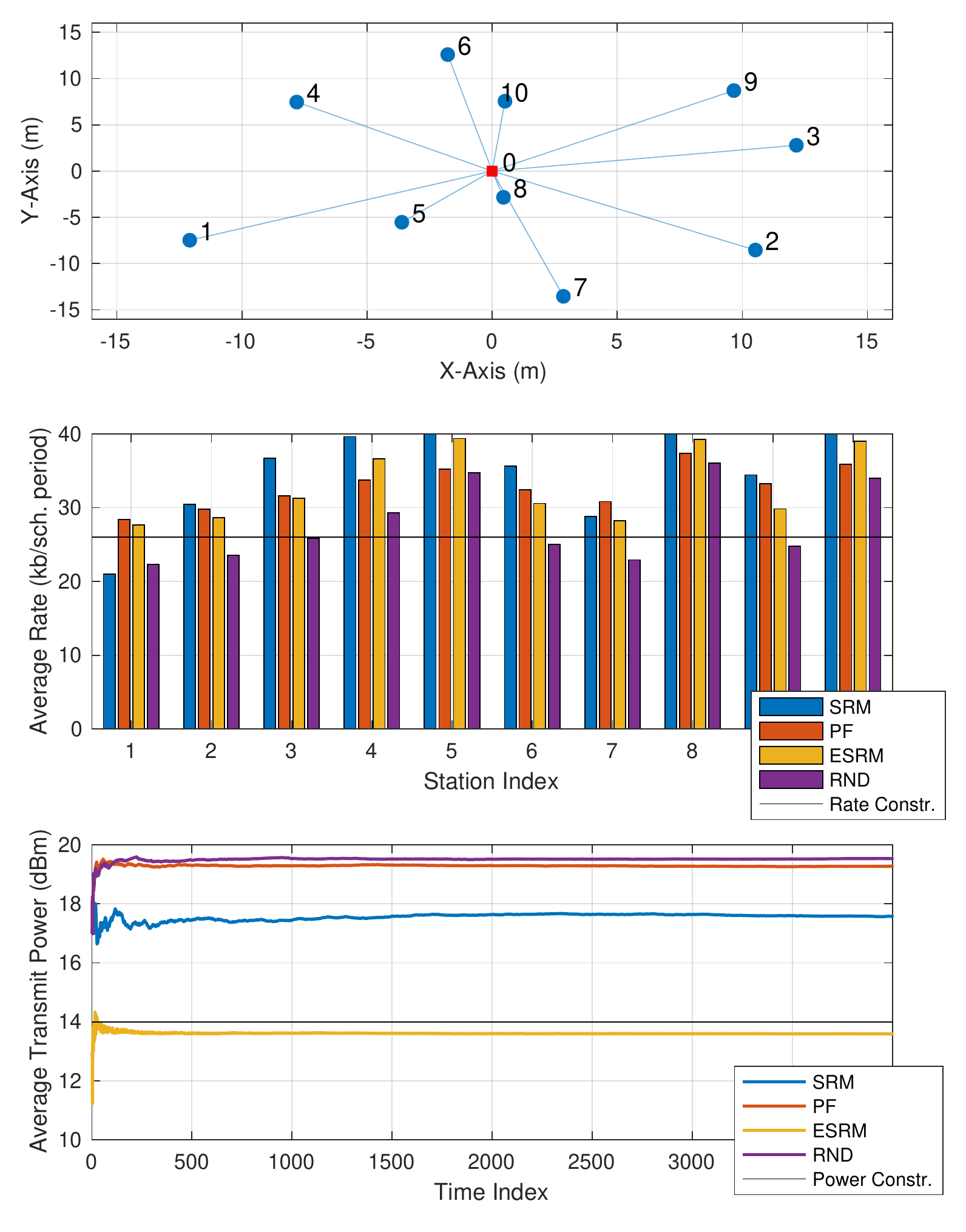}
		\caption{}
		\label{Fig:Result_3}
	\end{subfigure}
	\hfil
	\begin{subfigure}{1.\columnwidth}
		\includegraphics[width=\columnwidth]{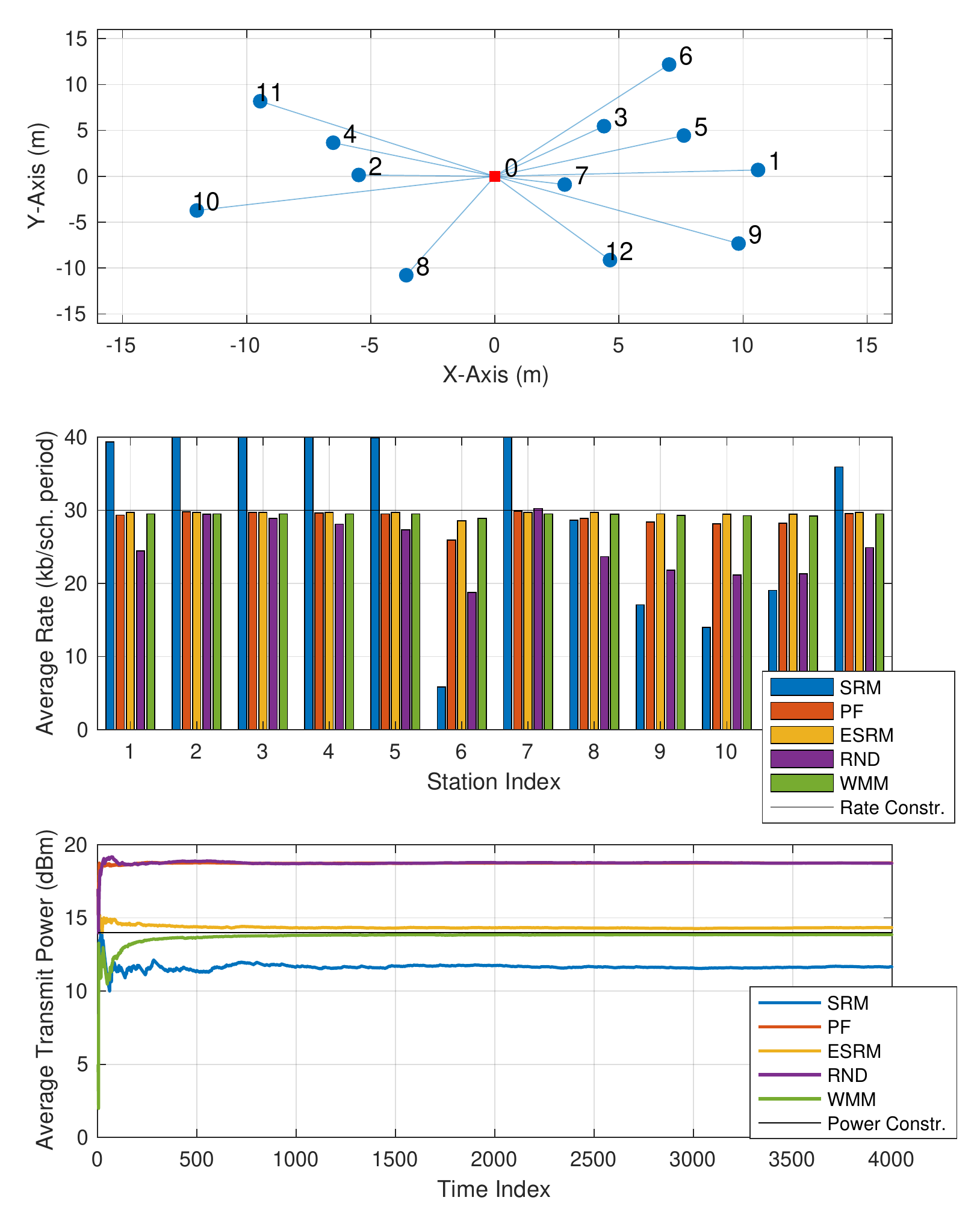}
		\caption{}
		\label{Fig:Result_4}
	\end{subfigure}
	\caption{Results for constrained utility maximization: (a) average rates in a network with $K=10$ stations, and average power consumption of the most distant station, i.e., $k=1$, of the network; (b) average rates in a network with $K=12$ stations, and average power consumption of the most distant station, i.e., $k=6$, of the network.}
\end{figure*}
\subsection{Results for Constrained Utility Maximization}
Consider the constraint maximization problem in \eqref{eq:stochastic_problem}. In order to assess the performance of the proposed framework, we fix the network topology and set the constraint values so that the problem is feasible. Specifically, we set the minimum average rate constraint to $r^{min} = 26$ kb/period and the maximum average power expenditure to $p^{max} = 14$ dBm for all stations. We implement ergodic sum-rate maximization (ESRM), which is given by the utility $U(\bar{\mathbf{r}}) = \sum_{k} \bar{r}_k$. Because $U(\bar{\mathbf{r}})$ is a linear function of the rates, we do not have to use auxiliary variables, and hence the variable $\phi_{k,n}(p_{k,n}(t))$ of DPP algorithm is given by
\begin{align*}
\phi_{k,n}(p_{k,n}(t)) = V r_{k,n}(t) &+ G_k(t)(r_{k,n}(t) - r^{min}) \\
&+ Q_k(t)(p^{max} - p_{k,n}(t)).
\end{align*}

In SRM, PF, and RND policies, all stations transmit at the maximum allowable power $P_{max} = 20$ dBm. In ESRM policy, stations select their transmit power from the set $\mathcal{P} = \{8,10,12,14,16,18,20\}$. We run the different policies for $4000$ scheduling periods, and calculate the average rates of stations as well as the moving average of the power consumption of the most distant station of the network. We consider only the power consumption of the most distant station since this station will consume the most power in order to satisfy the minimum average rate constraint of ESRM policy; the average power consumption of a station can be less than the target value because only a subset of sttations is selected to transmit in every scheduling period. The sum-rates achieved by SRM, ESRM, RND, and PF policies are $346, 330$, $278$, and $328$ kb/period, respectively. From  Fig. \ref{Fig:Result_3}, we see that ESRM policy guarantees minimum average rates and maximum average power expenditures for all stations at the price of a small reduction in the sum-rate (compared to SRM). Furthermore, ESRM and PF perform similarly, however ESRM attains significantly less average power expenditure. 

Finally, we study the case where the constrainted optimization problem is not feasible and employ the weighted max-min (WMM) fair algorithm of Section \ref{sec:wmm}. To make the problem infeasible, we increase the minimum average rate constraint to $30$ kb/period and the number of stations to $K=12$. In this case, the most distant station of the network, i.e., $k=6$, is expected to have the highest rate constraint violation. From Fig. \ref{Fig:Result_4}, we see that WMM policy indeed minimizes the maximum rate constraint violation while fulfilling the constraint on the average power expenditure. Moreover, the WMM performs similar to PF, but it does so with less average power consumption. The sum-rates achieved by SRM, ESRM, RND, PF, and WMM policies are $359, 354, 300, 347,$ and $352$ kb/period, respectively. 

\section{Conclusions}
Relying on the theory of Lyapunov optimization, we presented a novel approach for joint scheduling and resource allocation for IEEE 802.11ax UL OFDMA. A key merit of the proposed framework is that it can be used along with the TWT mechanism of IEEE 802.11ax to schedule stations in time and provide guarantees on their average power consumption and/or achievable rates whenever possible. Such a design approach is suitable for power-constrained IoT networks and battery-powered sensor networks. We finally showed through elaborate simulations that the derived resource allocation strategies outperfom state-of-the-art methods such as proportional fairness in terms of both average achievable rates and average power consumption of stations. In future work, it will be interesting to study the impact of the multiple RU patterns and the MIMO capabilities of IEEE 802.11ax. Another interesting direction of research is to condiser non- saturated traffic conditions, i.e., stations do not have always packets available for transmission, and study the scheduling and resource allocation problem subject to minimum average delay constraints.

\section*{Appendix A\\Proof of Lemma 1}
For each $\tau\in\{0,1,\dots\}$, it holds
\[
Q_k(\tau + 1) \geq Q_k(\tau) + p_k(\tau) - p^{max}_k.
\]
This is because $[x]^+ \geq x$. Thus, we have
\[
Q_k(\tau + 1) - Q_k(\tau) \geq p_k(\tau) - p^{max}_k. 
\]
Summing over $\tau\in\{0,1,\dots,t-1\}$ for some integer $t>0$ gives (through telescoping sum)
\[
Q_k(t) - Q_k(0) \geq \sum_{\tau=0}^{t-1}p_k(\tau) - tp^{max}_k.
\]
Diving by $t$ and using the fact that $Q_k(0) = 0, \forall k$, gives
\[
\frac{Q_k(t)}{t} \geq \frac{1}{t}\sum_{\tau=0}^{t-1}p_k(\tau) - p^{max}_k .
\]
Taking $\lim_{t\to\infty}\mathbb{E}[\cdot]$ and rearranging the terms, yields
\[
\bar{p}_k(t) \leq p^{max}_k  + \lim_{t\to\infty} \frac{\mathbb{E}[Q(t)]}{t}.
\]
Therefore if $Q_k$ is mean rate stable, then
\[
\bar{p}_k(t) \leq p^{max}_k.
\]

\section*{Appendix B\\Proof of  Bound on the Drift-Plus-Penalty}
We use the fact that $(\max[Q - b + a]^+)^2 \leq Q^2 + a^2 + b^2 + 2Q(a-b)$ for any $Q \geq 0$, $b\geq 0$, and $a \geq 0$. Then, squaring the update equation of queues $Q_k(t)$, $Z_k(t)$, and $G_k(t)$, we get the following inequalities

\begin{align}\label{ineq1}
Q_k^2(t+&1) = (\left[Q_k(t) - p^{max}_k  + p_k(t)\right]^+)^2 \nonumber\\ 
& \leq Q_k(t)^2 + p_k^2(t) + (p^{max}_k)^2 + 2Q_k(t)(p_k(t) - p^{max}_k)\nonumber\\
& \leq Q_k(t)^2 + P_{max}^2 + (p^{max}_k)^2 + 2Q_k(t)(p_k(t) - p^{max}_k)
\end{align}

\begin{align}\label{ineq2}
Z_k^2(t+&1)  = (\left[Z_k(t) - r_k(t) + \gamma_k(t)\right]^+)^2\ \nonumber \\
&\leq Z_k(t)^2 + r_k^2(t) + \gamma_k^2(t) + 2Z_k(t)(\gamma_k(t)-r_k(t))\nonumber\\
&\leq Z_k(t)^2 + 2R_{max}^2 + 2Z_k(t)(\gamma_k(t)-r_k(t))
\end{align}

\begin{align}\label{ineq3}
G_k^2(t+&1) = (\left[G_k(t) - r_k(t)+ r^{min}_{k}\right]^+)^2 \nonumber\\ 
& \leq G_k(t)^2 + r_k^2(t) + (r^{min}_k)^2 + 2G_k(t)(r^{min}_k-r_k(t))\nonumber\\
& \leq G_k(t)^2 + R_{max}^2 + (r^{min}_k)^2 + 2G_k(t)(r^{min}_k-r_k(t))
\end{align}
where inequalities \eqref{ineq1},  \eqref{ineq2}, and \eqref{ineq3} hold because $r_k(t)\leq R_{max}$, $\gamma_k(t)\leq R_{max}$, and $p_k(t) \leq P_{max}$, $\forall k, \forall t$. Using \eqref{ineq1},  \eqref{ineq2}, and \eqref{ineq3}, it is straightforward to derive the bound in \eqref{eq:drift}.

\section*{Appendix C\\Proof of Theorem 1}
Denote the drift-plus-penalty achieved by DPP as $\Delta(\bm{\Theta}(t)) - V\mathbb{E}[U(\bm{\gamma}_{\pi}(t)) | \bm{\Theta}(t)]$. Then \eqref{eq:drift} is written as
\begin{align}\label{eq:drift_expression}
\Delta(\bm{\Theta}(t)) - V\mathbb{E}[U(\bm{\gamma}_{\pi}(t)) |& \bm{\Theta}(t)] \leq B - V\mathbb{E}[U(\bm{\gamma}_{\omega}(t))]\ \nonumber\\
&+ \sum_{k=1}^K Q_k(t)\mathbb{E}[p_{k,\omega}(t) - p^{max}_k| \bm{\Theta}(t)] \nonumber\\ 
&+ \sum_{k=1}^K Z_k(t)\mathbb{E}[\gamma_{k,\omega}(t) - r_{k,\omega}(t) | \bm{\Theta}(t)] \nonumber \\
&+ \sum_{k=1}^K G_k(t)\mathbb{E}[r_{k}^{min} - r_{k,\omega}(t) | \bm{\Theta}(t)],
\end{align}
where $\omega$ refers to any other (possibly randomized) policy. In \cite{SNO1}, it was proven that there exists a randomized policy $\omega$ that satisfies for any $t\in\{0,1,2,\dots,\}$ the following inequalities:
\begin{align*}
- U(\bm{\gamma}_{\omega}(t)) &\leq - U^{opt} \\
\mathbb{E}[p_{k,\omega}(t) - p^{max}_k] &\leq 0, \quad \forall k \\
\mathbb{E}[\gamma_{k,\omega}(t) - r_{k,\omega}(t)] &\leq 0,  \quad \forall k \\
\mathbb{E}[r^{min}_{k} - r_{k,\omega}(t)] &\leq 0,  \quad \forall k.
\end{align*}
Plugging these inequalities into \eqref{eq:drift_expression} gives
\[
\Delta(\bm{\Theta}(t)) - V\mathbb{E}[U(\bm{\gamma}_{\pi}(t)) | \bm{\Theta}(t)] \leq B - VU^{opt}.
\]
By using iterated expectations and telescoping sums, we take
\begin{equation}\label{eq:bound_for_proof}
\mathbb{E}[L(\bm{\Theta}(t))]  - \mathbb{E}[L(\bm{\Theta}(0))] - V\sum_{\tau=0}^{t-1}\mathbb{E}[U(\bm{\gamma}_{\pi}(\tau))]  \leq tB - VtU^{opt}.
\end{equation}

\begin{enumerate}[(i)]
	\item \textit{Constraint Satisfaction}:
	Using the boundness assumption
	\[
	U_{\min} \leq \mathbb{E}[U(\bm{\gamma}_{\pi}(\tau))]  \leq U_{\max} , \quad \forall \tau \in\{0,1,\dots\},
	\]
	Eq. \eqref{eq:bound_for_proof} yields
	\begin{align*}
	\mathbb{E}[L(\bm{\Theta}(t))] \leq \mathbb{E}[L(\bm{\Theta}(0))] + Bt + Vt(U_{\max}-U^{opt})
	\end{align*}
	where 
	\begin{align*}
	\mathbb{E}[L(\bm{\Theta}(t))] = \frac{1}{2}\sum_{k=1}^K\mathbb{E}[Q^2_k(t)] &+ \frac{1}{2}\sum_{k=1}^K\mathbb{E}[Z^2_k(t)] \\
	&+ \frac{1}{2}\sum_{k=1}^K\mathbb{E}[G^2_k(t)].
	\end{align*}
	Therefore, the following inequalities hold for all $k$:
	\begin{align*}
	\mathbb{E}[Q^2_k(t)] &\leq 2\mathbb{E}[L(\bm{\Theta}(0))]  + 2Bt + 2Vt(U_{\max}-U^{opt}) \\
	\mathbb{E}[Z^2_k(t)] &\leq 2\mathbb{E}[L(\bm{\Theta}(0))]  + 2Bt + 2Vt(U_{\max}-U^{opt}) \\ 
	\mathbb{E}[G^2_k(t)] &\leq 2\mathbb{E}[L(\bm{\Theta}(0))]  + 2Bt + 2Vt(U_{\max}-U^{opt}).
	\end{align*}
	Because the variance ot $Q_k(t)$ cannot be negative, i.e., $	\mathbb{E}[Q^2_k(t)] \geq 	\mathbb{E}^2[Q_k(t)]$, it holds
	\[
	\mathbb{E}[Q_k(t)] \leq \sqrt{ 2\mathbb{E}[L(\bm{\Theta}(0))]  + 2Bt +2Vt(U_{\max}-U^{opt})}.
	\]
	Dividing by $t$, taking the limit $t\to\infty$ and using that $\mathbb{E}[L(\bm{\Theta}(0))] < \infty$, we finally get
	\[
	\lim_{t\to\infty} \frac{\mathbb{E}[Q(t)]}{t} = 0.
	\]
	Similarly, we have $\lim_{t\to\infty} \frac{\mathbb{E}[Z(t)]}{t} = 0$ and $\lim_{t\to\infty} \frac{\mathbb{E}[G(t)]}{t} = 0$; thus all queues are mean rate stable.
	
	\item \textit{Optimal Value}: Rearranging the terms, \eqref{eq:bound_for_proof}  yields 
	\begin{align*}\label{eq:ex_proof}
	VtU^{opt} - \mathbb{E}[L(\bm{\Theta}(0))] - tB  &\leq VtU^{opt}  \\
	&+ \mathbb{E}[L(\bm{\Theta}(t))] - \mathbb{E}[L(\bm{\Theta}(0))] \\
	&- tB \\
	& \leq  V\sum_{\tau=0}^{t-1}\mathbb{E}[U(\bm{\gamma}_{\pi}(\tau))].
	\end{align*}
	Therefore, we have 
	\[
	V\sum_{\tau=0}^{t-1}\mathbb{E}[U(\bm{\gamma}_{\pi}(\tau))] \geq VtU^{opt} - \mathbb{E}[L(\bm{\Theta}(0))] - tB.
	\]
	Dividing by tV and taking the limit $t\to\infty$ yields
	\[
	\lim_{t\to\infty}\frac{1}{t}\sum_{\tau=0}^{t-1}\mathbb{E}[U(\bm{\gamma}_{\pi}(\tau))] \geq U^{opt} -\frac{B}{V} 
	\]
	or equivalently
	\[
	\overline{U(\bm{\gamma}_{\pi})} \geq U^{opt} -B\epsilon
	\]
	where $\epsilon = 1/V$. Since all virtual queues are mean rate stable,  \eqref{eq:inequalities} holds, and hence
	\[
	U(\bar{\mathbf{r}}_{\pi}) \geq U^{opt} - O(\epsilon).
	\]
\end{enumerate}

\section*{Appendix D\\Simulation Parameters}

\begin{table}[H]
	\centering
	\caption{\ac{MCS} for the 20MHz channel and \ac{RUs} \\of $24$ data subcarriers \cite{802.11ax_Draft}.}
	\begin{tabular}{c | c | c } 
		\hline
		\textbf{Index} & \textbf{MCS} & \textbf{Minimum SNR (dBm)} \\ 
		\hline
		1 &BPSK, $1/2$  		 & $-82$\\
		2 &QPSK, $1/2$  		& $-79$\\
		3 &QPSK, $3/4$  		& $-77$\\
		4 &16-QAM, $1/2$    & $-74$\\
		5 &16-QAM, $3/4$    & $-70$\\
		6 &64-QAM, $2/3$    & $-66$\\
		7 &64-QAM, $3/4$    & $-65$\\
		8 &64-QAM, $5/6$    & $-64$\\
		9 &256-QAM, $3/4$  & $-59$\\
		10 &256-QAM, $5/6$  & $-57$\\
		\hline
	\end{tabular}
	\label{table: MCS}
\end{table}

\section*{Acknowledgment}
This work has been partially supported by a Gift from the Cisco University Research Program (CG\#$890107$, Towards Deterministic Channel Access in High-Density WLANs) Fund, a corporate advised fund of Silicon Valley Community Foundation, and by the Catalan Government SGR grant for research support (2017-SGR-1188).

\end{document}